\documentclass[floats,floatfix,showpacs,amssymb,prd,onecolumn,superscriptaddress,nofootinbib,twocolumn]{revtex4-2}

\usepackage{graphicx}
\usepackage{dcolumn}
\usepackage{bm}
\usepackage{amsmath}
\usepackage[colorlinks=true,linkcolor=blue,citecolor=blue,urlcolor=blue]{hyperref}
\usepackage{dsfont}

\usepackage{xcolor}
\usepackage{ulem}
\usepackage{makecell}

\newcommand{\bfx}{{\bf x}}

\newcommand\comment[1]{}
\newcommand{\Si}{\rm{Si}}

\begin{document}

\title{Rapid and accurate numerical evolution of linear cosmological perturbations \\with non-cold relics} 

\author{Nanoom Lee}
\email{nanoom.lee@jhu.edu}
\affiliation{William H. Miller III Department of Physics \& Astronomy, Johns Hopkins University, Baltimore, Maryland 21218, USA}
\author{José Luis Bernal}
\email{jlbernal@ifca.unican.es}
\affiliation{Instituto de Física de Cantabria (IFCA), CSIC-Univ. de Cantabria, Avda. de los Castros s/n, E-39005 Santander, Spain}
\author{Sven G\"unther}
\affiliation{Institute for Theoretical Particle Physics and Cosmology (TTK), RWTH Aachen University, D-52056 Aachen, Germany}
\author{Lingyuan Ji}
\affiliation{Department of Physics, University of California, Berkeley, 366 Physics North MC 7300, Berkeley, California 94720, USA}
\author{Marc Kamionkowski}
\email{kamion@jhu.edu}
\affiliation{William H. Miller III Department of Physics \& Astronomy, Johns Hopkins University, Baltimore, Maryland 21218, USA}

\begin{abstract}
We describe the implementation of a new approach to the numerical evaluation of the effects of non-cold relics on the evolution of cosmological perturbations. The Boltzmann hierarchies used to compute the contributions of these relics to the stress-energy tensor are replaced with a set of integral equations. These integral equations take the form of convolutions and are solved iteratively with the rest of the system. We develop efficient algorithms for evaluating these convolutions using non-uniform fast Fourier transforms (NUFFTs). This approach enables efficient and accurate evaluation of the cosmic microwave background anisotropies and matter power spectra, all the way through the history of the Universe, without relying on semi-analytic approximations at late times. We implement this method in the Boltzmann solver \texttt{CLASS}, resulting in a new code called \texttt{CLASSIER} (for \texttt{CLASS} Integral Equation Revision), and apply it to massive-neutrino perturbations as a demonstration. The implementation is optimized to accurately capture the distinct behaviors of perturbations in both super-/near-horizon and sub-horizon regimes. Our results match the accuracy of a fully converged Boltzmann hierarchy solution while avoiding numerical artifacts from truncation of the Boltzmann hierarchy at finite multipole and offering substantial speedups depending on the required precision and the range of scales of interest. This new framework provides a practical and robust alternative for the truncated Boltzmann hierarchy approach, especially for studying beyond $\Lambda$CDM non-cold relics with signatures on small scales. \texttt{CLASSIER} is publicly available at \href{https://github.com/nanoomlee/CLASSIER}{https://github.com/nanoomlee/CLASSIER}.
\end{abstract}

\maketitle

\section{Introduction}

Accurate modeling of cosmological perturbations is essential for extracting reliable information about the Universe from present and future cosmological surveys. With upcoming and ongoing high-precision observations such as those from CMB-S4 \cite{CMB-S4:2016ple, Abazajian:2019eic, Abazajian:2022nyh}, Simons Observatory \cite{SimonsObservatory:2018koc,SimonsObservatory:2019qwx}, CMB-HD \cite{Sehgal:2019ewc,CMB-HD:2022bsz}, DESI \cite{DESI:2016fyo,DESI:2019jxc}, the Rubin Observatory Legacy Survey of Space and Time (LSST) \cite{LSST:2008ijt,LSSTDarkEnergyScience:2012kar, LSSTDarkEnergyScience:2018jkl}, the Nancy Grace Roman Space telescope \cite{Spergel:2015sza}, and the Euclid \cite{EUCLID:2011zbd}, even subtle effects in the evolution of perturbations of each species must be modeled with care. Among these, massive neutrinos and more general exotic non-cold dark-matter (NCDM) relics play a small but key role: their free-streaming behavior suppresses the growth of structure on small scales, modifies the evolution of metric perturbations, and leaves distinct imprints in both the cosmic microwave background (CMB) and large-scale structure~\cite{Lesgourgues:2006nd}.

The conventional approach for computing perturbations of most species, including non-cold relics, involves solving a Boltzmann hierarchy for the momentum-dependent multipole moments of the perturbed phase-space distribution function. For practical purposes, this hierarchy must be truncated at a finite maximum multipole $\ell_{\text{max}}^{\rm NCDM}$, and numerical results are sensitive to this choice, with numerical inaccuracies potentially arising~\cite{Ma:1995ey, class}. Ref.~\cite{Lesgourgues:2011rh} developed a powerful ultra-relativistic fluid approximation for the late-time evolution of massive-neutrino perturbations.  This semi-analytic treatment is, however, ultimately calibrated to precise Boltzmann results for massive neutrinos.  It is thus limited in precision and cannot be applied to other non-cold relics or used to study massive-neutrino models with nontrivial phase-space distributions.

Multipole moments can also be computed with the so-called line-of-sight integral, as suggested for CMB anisotropies~\cite{Zaldarriaga:1995gi,Seljak:1996is} and first implemented in \texttt{CMBfast} \cite{Seljak:1996is}, which became the numerical scheme that the modern Boltzmann solvers such as \texttt{CLASS}~\cite{class, Lesgourgues:2011rh} and \texttt{CAMB}~\cite{Lewis:1999bs, Howlett:2012mh} are based on. On the other hand, it was pointed out that this integral-equation (IE) approach could entirely replace the infinite Boltzmann hierarchy \cite{Weinberg:2003ur, Baskaran:2006qs,Weinberg:2006hh}. This approach was applied for primordial tensor perturbations \cite{Pritchard:2004qp,Flauger:2007es,Loverde:2022wih}, and suggested for photon and massive neutrino perturbations using an iterative algorithm~\cite{Kamionkowski:2021njk, Ji:2022iji}.

Being collisionless, NCDM species are a special case. The multipole moments of their perturbed distribution function can be expressed as a convolution of known analytic kernels with gravitational source functions. This convolution structure makes the numerical implementation efficient, as the integrals can be computed using fast Fourier transforms \cite{Ji:2022iji}. Since the metric perturbations in the source terms depend on moments of the distribution function, the system is solved iteratively until convergence. While early results confirmed the accuracy of the IE method at early times, its viability and efficiency as a full replacement for the Boltzmann hierarchy throughout entire cosmic history has not yet been fully explored. Building on these earlier works, we implement the IE approach into the public Boltzmann solver \texttt{CLASS} \cite{class,Lesgourgues:2011rh} for collisionless non-cold relics and apply it to massive neutrinos as a numerical demonstration.

Our efforts result in a modification of \texttt{CLASS} that we call \texttt{CLASS} Integral Equation Revision (\texttt{CLASSIER}).\footnote{Available at \href{https://github.com/nanoomlee/CLASSIER}{https://github.com/nanoomlee/CLASSIER}.} We validate the accuracy of our implementation for the matter power spectrum today against the results of \texttt{CLASS}, using the Boltzmann hierarchy with a cut-off $\ell_{\rm max}$ high enough to ensure convergence, up to  $k\sim 100\;{\rm Mpc}^{-1}$.\footnote{This does not necessarily imply a breakdown of accuracy at $k>100\;{\rm Mpc}^{-1}$.} Furthermore, we find that our implementation is several times faster than solving the Boltzmann hierarchy for accurate calculations at small scales.

The rest of this paper is organized as follows. In Section \ref{sec:motivation}, we first elaborate on the motivation behind this work. Section \ref{sec:integral-sol} reviews the integral-equation formulation for collisionless particles. In Section \ref{sec:convolution}, we discuss the evaluation of the convolution integrals, describe our numerical implementation, and present convergence tests for our iterative scheme. Section \ref{sec:comparison} compares the accuracy and performance of our approach and the truncated Boltzmann hierarchy approach. We conclude in Section \ref{sec:conclusion} with a discussion of implications and possible extensions. We implement our approach on the public \texttt{CLASS v3.2.2}, and show results assuming the default cosmological parameters with one massive neutrino species of mass $0.06\;{\rm eV}$, unless stated otherwise.

\begin{figure*}[t!]
\includegraphics[width = \linewidth,trim= 10 20 10 0]{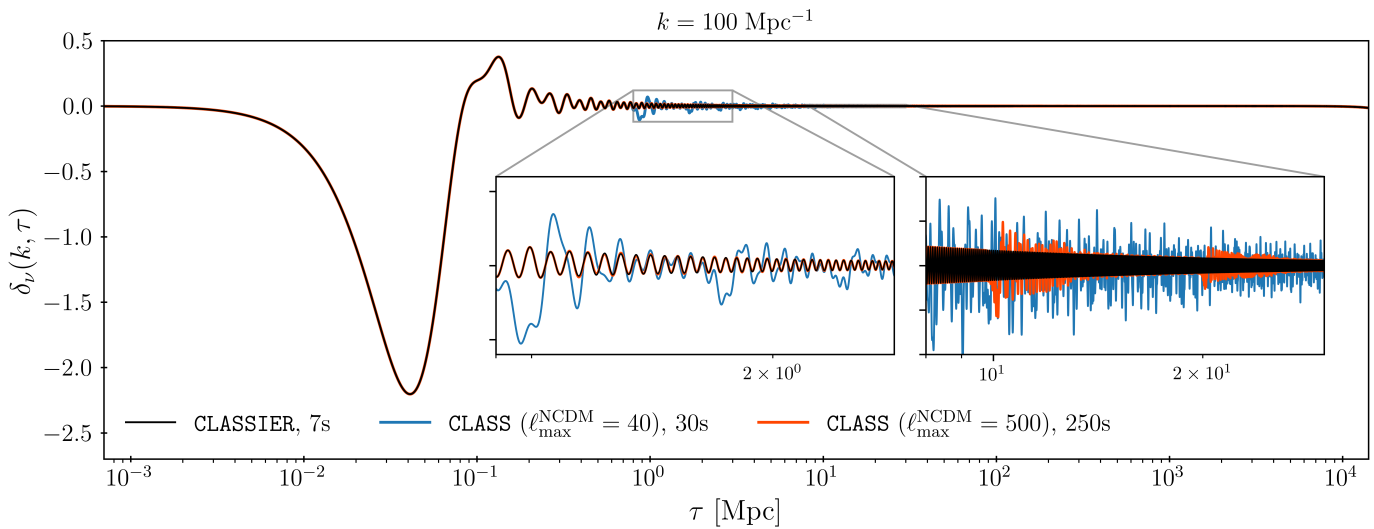}
\caption{Massive neutrino density perturbations $\delta_\nu(k,\tau)$ for the $k=100\;{\rm Mpc}^{-1}$ mode from three different calculations. \texttt{CLASSIER} (black), which is presented in this work, solves the integral equations with an iterative approach, which replaces the usual Boltzmann hierarchy for massive neutrinos. Our result shown here is obtained with a single iteration. \texttt{CLASS} (blue and orange) solves the Boltzmann hierarchy truncated at $\ell_{\rm max}^{\rm NCDM}=40$ and 500, respectively, for massive neutrinos. The truncation induces numerical errors in $\delta_\nu$ starting at $\tau \sim 2\ell_{\rm max}^{\rm NCDM}/k$. These errors can be seen in two inset plots: the left shown for $0.8\;{\rm Mpc}\leq \tau \leq 3\;{\rm Mpc}$ and the right for $8\;{\rm Mpc}\leq \tau \leq 30\;{\rm Mpc}$. We also report the time in seconds taken to solve the coupled system of cosmological perturbations of $k=100\;{\rm Mpc}^{-1}$ mode, including 0th iteration in case of \texttt{CLASSIER} which is with $\ell_{\rm max}^{\rm NCDM}=17$ and the fluid approximation for massive neutrino turned on at $\tau=31/k$.}
\label{fig:delta_nu-k100}
\end{figure*}

\section{Motivation -- Limitations of\\Boltzmann hierarchy approach}
\label{sec:motivation}
In linear cosmological perturbation theory, the Boltzmann equation of NCDM particles depends on the amplitude of the particle momentum and the angle between the momentum and the Fourier wavevector $\vec{k}$ \cite{Ma:1995ey}.\footnote{The case for massless particles is simpler, since the dependence on the norm of the momentum can be factored out.} It is typically rewritten as an infinite set of coupled differential equations for the momentum-dependent multipole moments of the perturbed phase-space-density distribution function. This system of coupled differential equations is referred to as the Boltzmann hierarchy. Even though truncation methods at finite $\ell_{\rm max}$ multipoles~\cite{Ma:1995ey} have been developed, this results in a total of $N_{\rm NCDM}\times \ell_{\rm max}\times N_q$ differential equations added to the system to solve, where $N_q$ is the number of momentum bins used and $N_{\rm NCDM}$ is the number of NCDM species considered. As expected, this leads to a substantial computational cost.

To overcome such computational difficulties, fluid approximations for massive neutrino perturbations have been developed \cite{Hu:1998kj,Lewis:2002nc,Shoji:2010hm,Lesgourgues:2011rh,Nascimento:2023psl}. Ref.~\cite{Lesgourgues:2011rh} provides an impressively accurate prediction for observables like the matter power spectrum, with errors $\lesssim \mathcal{O}(0.1\%)$ at $z=0$, only requiring solving the Boltzmann hierarchy up to a certain early epoch. Despite this, small but non-negligible numerical errors can still appear at early times, arising from truncating the hierarchy, as previously noted in Ref.~\cite{Ma:1995ey,Lesgourgues:2011rh}. Besides, the approximation introduces large $\mathcal{O}(1)$ errors in massive neutrino density perturbations themselves, especially at late times. These discrepancies can become problematic when accurately computing observables that are sensitive to the evolution of non-cold species. Furthermore, the development of a fluid approximation is often model-specific and may not generalize easily to exotic DM models (e.g., see Ref.~\cite{FrancoAbellan:2020xnr} for the fluid approximation for their warm dark matter model).

Nonetheless, solving the Boltzmann hierarchy up to a large $\ell_{\rm max}^{\rm NCDM}$ for all times is necessary for high-precision predictions. For instance, we have verified that convergence in the small-scale matter power spectrum today is achieved for $\ell_{\rm max}^{\rm NCDM}\sim500$. Even so, there are numerical artifacts in the neutrino perturbations due to the truncation of the Boltzmann hierarchy.

In summary, while the development of the Boltzmann hierarchy and its numerical solvers is a major success in modern cosmology, limitations persist: 1) truncation-induced numerical artifacts in the perturbation evolution, 2) model-specific fluid approximation that may be inaccurate at late times, and 3) long runtimes when using high $\ell_{\rm max}^{\rm NCDM}$ values to ensure precision (exacerbated for high $k$). Instead, the IE can overcome these challenges. To illustrate this, we show the evolution of massive neutrino density perturbations for $k = 100\;{\rm Mpc}^{-1}$ using the Boltzmann hierarchy with $\ell_{\rm max}^{\rm NCDM} = 40$ (blue) and $500$ (orange), and compare them with the results of the IEs (black) in Fig.~\ref{fig:delta_nu-k100}. The onset of truncation-induced errors is visible at conformal times $\tau \sim 2\ell_{\rm max}^{\rm NCDM}/k$, consistent with expectations. Note that these numerical artifacts do not appear when using the IEs. Furthermore, the runtime for evolving perturbations of $k=100\;{\rm Mpc}^{-1}$ mode is reduced by a factor of $\sim4$ (40) compared to the hierarchy method with $\ell_{\rm max}^{\rm NCDM}=40$ (500). This demonstrates how effective our approach is in terms of accuracy and performance compared to the full Boltzmann hierarchy approach. 

\newpage

\section{Integral solutions of\\collisionless Boltzmann equation}
\label{sec:integral-sol}

The Fourier-space, linearized Boltzmann equation for NCDM scalar perturbations in synchronous gauge is given by~\cite{Ma:1995ey}
\begin{eqnarray}
\frac{\partial \Psi}{\partial \tau}+ik\mu\frac{q}{\epsilon}\Psi + \frac{d\ln f_0}{d\ln q}\left[ \dot{\eta} - \frac{\dot{h}+6\dot{\eta}}{2}\mu^2\right] = 0,
\label{eq:Boltzmann}
\end{eqnarray}
where $\Psi$ is the fractional perturbation of the phase-space-density function $f(\vec{q},\vec{k},\tau) = f_0(q) [1+\Psi(q,k,\mu,\tau)]$ with comoving momentum $\vec{q}$ and energy $\epsilon(q,\tau) = \sqrt{q^2+a(\tau)^2m^2/T_0^2}$, both in terms of the NCDM temperature $T_0$ today, and particle mass $m$.\footnote{These definitions of $q$ and $\epsilon$ follow that of Ref.~\cite{Lesgourgues:2011rh}, not Ref.~\cite{Ma:1995ey}.} Due to symmetry \cite{Ma:1995ey}, $\Psi$ depends only on $q\equiv |\vec{q}|$, the Fourier wavenumber $k\equiv |\vec{k}|$, and the angle formed by these two vectors $\mu=\hat{k}\cdot\hat{q}$. The variables $h(k,\tau)$ and  $\eta(k,\tau)$ are the synchronous gauge metric perturbations, and the overdots denote derivative with respect to conformal time $\tau$.\footnote{Note that this notation for derivatives is different from that of previous studies, Refs.~\cite{Kamionkowski:2021njk,Ji:2022iji}.}

Integrating Eq.~\eqref{eq:Boltzmann} along the particle trajectory, we find a formal solution in integral form
\begin{eqnarray}
\Psi(\tau_f) &=& e^{-i\mu k \chi_q(\tau_i,\tau_f)}\Psi(\tau_i) \nonumber\\
&+& \int_{\tau_i}^{\tau_f} e^{-i\mu k \chi_q(\tau,\tau_f)} \left[ -\dot{\eta} + \frac{\dot{h}+6\dot{\eta}}{2}\mu^2\right] \frac{d\ln f_0}{d\ln q} d\tau,~~~~~
\end{eqnarray}
where the comoving horizon $\chi_q(\tau_1,\tau_2)\equiv \int_{\tau_1}^{\tau_2} d\tau' q/\epsilon(\tau')$ accounts for the time-varying relation between comoving momentum $q$ and energy $\epsilon$. We define the multipole moments $\Psi_\ell \equiv (i^\ell/2)\int_{-1}^1 \Psi(\mu) P_\ell(\mu)$ where $P_\ell$ are the Legendre polynomials. Then, using the identity for the spherical Bessel function
\begin{eqnarray}
\frac{d^n}{dx^n}j_\ell(x) = \frac{i^\ell}{2}\int_{-1}^1 e^{-i \mu x}(-i \mu)^n P_\ell(\mu)d\mu,
\end{eqnarray}
we obtain
\begin{eqnarray}
     \Psi_\ell(\tau_f)& = & \sum_{\ell'=0}^\infty (-1)^{\ell'} (2\ell'+1) W_{\ell\ell'}[k\chi_q(\tau_i,\tau_f)]\Psi_{\ell'}(\tau_i)\nonumber \\
    &&+ \int_{\tau_i}^{\tau_f} \frac{d \ln f_0}{d \ln q} \, d\tau \, \nonumber \\
   &\times& \bigg\{ 
    -j_\ell[k\chi_q(\tau, \tau_f)]\; \dot{\eta} - j''_\ell[k\chi_q(\tau, \tau_f)] \;\frac{\dot{h} + 6\dot{\eta}}{2}
    \bigg\},~~~~
    \label{eq:Psi_ell}
\end{eqnarray}
where $j''_\ell(x) \equiv d^2 j_\ell(x) / dx^2$ and
\begin{eqnarray}
W_{\ell\ell'}(x) &\equiv& \frac{i^{\ell+\ell'}}{2} \int_{-1}^1 e^{-i\mu x}P_\ell(\mu)P_{\ell'}(\mu) d\mu \nonumber\\
&=& i^{\ell'} P_{\ell'}\left(i\frac{d}{dx}\right)j_\ell(x).
\end{eqnarray}
We choose the initial time $\tau_i$ sufficiently early so that $\Psi_\ell(\tau_i)=0$ for $\ell>2$. The second integral term can be rewritten as a convolution, as we describe in the following Section \ref{sec:convolution}.

From these moments one can compute the contributions of non-cold relics to the matter-density, pressure perturbations, the peculiar velocity, and anisotropic stress as
\begin{eqnarray}
\delta \rho &=& 4 \pi a^{-4} T_0^4 \int q^2 dq \, \epsilon \, f_0(q T_0) \Psi_0(q), \nonumber\\
\delta P &= &\frac{4 \pi}{3} a^{-4} T_0^4 \int q^2 dq \, \frac{q^2}{\epsilon} \, f_0(q T_0) \Psi_0(q),  \nonumber\\
(\bar{\rho} + \bar{P}) \theta &=& 4 \pi k a^{-4} T_0^4 \int q^2 dq \, q f_0(q T_0) \Psi_1(q),  \nonumber\\
(\bar{\rho} + \bar{P}) \sigma &=& \frac{8 \pi}{3} a^{-4} T_0^4 \int q^2 dq \, \frac{q^2}{\epsilon} \, f_0(q T_0) \Psi_2(q),
\label{eq:neutrino-perturb}
\end{eqnarray}
respectively. 
These quantities are all the sources to the metric-perturbation variables appearing in the perturbed Einstein's equations. Hence, we only need to compute Eq.~\eqref{eq:Psi_ell} for $\ell \leq2$.\footnote{In principle, one could compute only the $\ell=0$ or $\ell=2$ moment and use it to solve for the rest of two multipole moments through their evolution equations [e.g., see Eq.~(56) in Ref.~\cite{Ma:1995ey}] together with the rest of system. However, this approach is numerically inefficient as it requires extremely high resolution in the integral solution not to induce artificial numerical errors in the other two multipole moments which can propagate to and be amplified at late time.} Since those same metric perturbations enter as sources in the integral solutions above, this coupled system must be solved iteratively to achieve self-consistency. 

We implement this iterative IE approach in \texttt{CLASS} as follows. We first obtain the metric perturbations ($\dot{h}$ and $\dot{\eta}$) by solving the standard Boltzmann hierarchy with a standard $\ell_{\rm max}^{\rm NCDM}=17$ and switching the fluid approximation on as usual. We refer to this initial computation, which does not yet involve the integral solutions, as the 0th iteration. The resulting metric perturbations are fed to the integral equations, the solutions of which are used to solve the evolution equations for the metric and all other species until everything converges after $N_{\rm iter}$ iterations.

We show results for an applications for a massive neutrino, using for the 0th iteration the default  $\ell_{\rm max}^{\rm NCDM}=17$ and $\tau = 31/k$ as the time to switch on the fluid approximation. Because the truncated Boltzmann hierarchy provides accurate results up to $\tau \lesssim \ell_{\rm max}^{\rm NCDM}/k$ \cite{Lesgourgues:2011rh}, we retain the 0th iteration results for $\tau \leq \ell_{\rm max}^{\rm NCDM}/k$ in all subsequent iterations, and apply the integral solution only for $\tau > \ell_{\rm max}^{\rm NCDM}/k$.\footnote{This automatically defines the range of $k$ modes to which we apply our implementation: $k>\ell_{\rm max}^{\rm NCDM}/\tau_0$, where $\ell_{\rm max}^{\rm NCDM}=17$ by default and $\tau_0$ is the conformal time today.}

\section{Evaluation of the convolutions}
\label{sec:convolution}

\subsection{Setting up the convolution structure}
\label{sec:setting}

\begin{table*}[!ht]
    \centering
    \begin{tabular}{c|c}
        $K(\xi)$ & $\hat K(\omega)$ \\ \hline
        $j_0(\xi)$ & $-\frac{i}{2}(C_{\text{Ei}} - C_{\text{Log}})$ \\
        $j_1(\xi)$ & $\frac{\omega}{2}(C_{\text{Ei}} - C_{\text{Log}}) + \frac{i}{2 \xi_0} C_{\text{Exp}} + 1$ \\
        $j_2(\xi)$ & $\frac{1}{4} i \left(3 \omega^2-1\right) (C_{\text{Ei}} - C_{\text{Log}}) + \frac{-3 \omega \xi_0+3 \xi_0+3 i}{4 \xi_0^2} C_{\text{Exp}} + \frac{3}{2 \xi_0} e^{i (\omega-1) \xi_0} + \frac{3 i \omega}{2}$ \\ \hline
        $j''_l(\xi)$ & $[j'(\xi) e^{i\omega\xi}]^{\xi_0}_0 - (i\omega)[j(\xi) e^{i\omega\xi}]^{\xi_0}_0 + (i\omega)^2 \mathcal F[j_l](\omega)$
    \end{tabular}
    \caption{The kernel functions $K(\xi)$ and their Fourier transforms $\hat K(\omega)$ with $K(\xi)$ restricted within $0\leq\xi\leq\xi_0$ [Eq.~\eqref{eq:Khat}].  To make the expressions succinct, we have adopted the following: $C_{\text{Ei}} \equiv \text{Ei}[i(\omega+1)\xi_0] - \text{Ei}[i(\omega-1)\xi_0]$, $C_{\text{Log}} \equiv \ln [(\omega+1)/(\omega-1)]$, and $C_{\text{Exp}} \equiv \exp[i(\omega+1)\xi_0] - \exp[i(\omega-1)\xi_0]$, where Ei implies the exponential integral.  Note that, when $\omega < 1$, the logarithm in $C_{\text{Log}}$ will be evaluated on the negative real axis, which is a branch cut.  We define $\ln z \equiv \ln |z| + i \arg z$ for $-\pi < \arg z \leq \pi$, thus allowing evaluation on the branch cut.  Also note that $\omega = 1$ is a removable singularity, so the formulas provided will work as long as $\omega = 1$ is not directly evaluated in numerical work.}
    \label{tab:kernel-fourier-transform-short-interval}
\end{table*}

We evaluate the integrals in Eq.~\eqref{eq:Psi_ell} for $\ell=0,1,2$. Before getting into technical details, let us think about what these integrals mean.  The quantity $\Psi_0(\tau,q)$ is the perturbation to the neutrino phase-space density at the origin at conformal time $\tau$ for neutrinos of (dimensionless) momentum $q$ induced by a primordial curvature perturbation with spatial dependence $\zeta(\bfx) \propto j_0(kr)$, where $r=|\bfx|$. The integral sums the contributions to the neutrino phase-space density that arise over the trajectory of the neutrino.  It therefore receives contributions all the way to comoving free-streaming length of a particle of momentum $q$.  For $\ell=1$, it is the same, except for a curvature perturbation with spatial dependence $j_1(kr)\cos\theta$, where $\theta$ is the polar angle.  And for $\ell=2$, it is $j_2(kr)P_2(\cos\theta)$, where $P_\ell(x)$ is the Legendre polynomial.

Back to the technical details: after the change to variables $\xi(k,q,\tau) = k \chi_q(\tau_i,\tau)$ in Eq.~\eqref{eq:Psi_ell}, we have integrals of the form,
\begin{equation}
     I(\xi) = \int_0^{\xi}\, d\xi'\,G(\xi') K(\xi-\xi'),
\label{eq:I_xi}
\end{equation}
for values of $\xi$ in the range $0\leq\xi\leq\xi_0 =\xi(k,q,\tau_0)$, where $G$ and $K$ are defined as either
\begin{eqnarray}
G(\xi) &\equiv&
\begin{cases}
-\dot{\eta}\frac{\epsilon}{qk}\frac{d\ln f_0}{d\ln q} & {\rm for} \quad 0 \leq\xi\leq\xi_0, \\
0 & {\rm otherwise},
\end{cases} \\
\quad K(\xi) &\equiv&
\begin{cases}
j_\ell(\xi) & {\rm for} \quad 0 \leq\xi\leq\xi_0, \\
0 & {\rm otherwise},
\end{cases}
\label{eq:GA}
\end{eqnarray}
or
\begin{eqnarray}
G(\xi) &\equiv&
\begin{cases}
 -\frac{\dot{h}+6\dot{\eta}}{2}\frac{\epsilon}{qk}\frac{d\ln f_0}{d\ln q} & {\rm for} \quad 0 \leq\xi\leq\xi_0, \\
 0 & {\rm otherwise},
 \end{cases}\\
 K(\xi) &\equiv&
 \begin{cases}
  j''_\ell(\xi) & {\rm for} \quad 0 \leq\xi\leq\xi_0, \\
  0 & {\rm otherwise},
\end{cases}
\label{eq:GB}
\end{eqnarray}
where we omit $k$, $q$, and $\xi$ dependences for simplicity. Note that $G(\xi)$ is meaningfully defined only within $0\leq\xi\leq\xi_0$, and this in turn makes the contribution from $K(\xi)$ restricted to $0\leq\xi\leq\xi_0$.\footnote{Note that there is no restriction for the kernels $K(\xi)$ to vanish for $\xi>\xi_0$. But this definition is beneficial for our purpose of numerical implementation since the Fourier transforms of spherical Bessel functions (and its derivatives) supported on $[0,+\infty)$ have singularities at $\omega=\pm1$, where $\omega$ is the Fourier frequencies, and those singularities are avoidable when the kernel functions are supported on a finite range, as we define here.} Given that $\tau_0\sim 10^4$ Mpc, $\xi_0 \lesssim 10^4\, k\, {\rm Mpc}$, which can get as high as $10^6$ for $k\sim 100\, {\rm Mpc}^{-1}$.  For smaller $q$, $\xi_0$ is reduced by $ \chi_q(\tau_i,\tau_0)/\tau_0$, the ratio of the comoving free-streaming horizon $\chi_q(\tau_i,\tau_0)$ and $\tau_0$.

We then rewrite the integral form in Eq.~\eqref{eq:I_xi} as a convolution over all $\xi$ values:
\begin{equation}
     I(\xi) = \int_{-\infty}^\infty\, d\xi'\,G(\xi') K(\xi-\xi').
\label{eq:convolution}
\end{equation}
To evaluate $I(\xi)$, we first perform Fourier transforms of $K(\xi)$ and $G(\xi)$ of which the integration interval can be limited as,
\begin{eqnarray}
    \hat G(\omega) &=& \int_{-\infty}^\infty \, d\xi\, e^{i \omega \xi} G(\xi) = \int_0^{\xi_0} \, d\xi\, e^{i \omega \xi} G(\xi),\label{eq:Ghat}\\
    \hat K(\omega) &= &\int_{-\infty}^\infty \, d\xi\, e^{i \omega \xi} K(\xi) = \int_0^{\xi_0} \, d\xi\, e^{i \omega \xi} K(\xi)\,,
    \label{eq:Khat}
\end{eqnarray}
and then multiply those Fourier transforms and perform the inverse Fourier transforms back to configuration space:
\begin{eqnarray}
    I(\xi) = \int_{-\infty}^\infty\, \frac{d\omega}{2\pi}\, e^{- i\omega\xi} \,\hat G(\omega) \hat K(\omega).
\end{eqnarray}

\begin{table*}[]
    \centering
    \begin{tabular}{c|c|c|c|c}
         $K(\xi)$ & $\int K(\xi) d\xi$ & $\int\, K(\xi) \xi d\xi$ & $\int\, K(\xi) \xi^2 d\xi$ & $\int\, K(\xi)\xi^3 d\xi$ \\ \hline
         $j_0(\xi)$ & Si$(\xi)$ & $-\cos\xi$ & $-\xi\cos\xi + \sin\xi$  & $-(\xi^2-2)\cos\xi + 2\xi\sin\xi$ \\
         $j_1(\xi)$ & $-\frac{\sin\xi}{\xi}$ & $-\sin\xi + \Si(\xi) $ & $-2\cos\xi -\xi\sin\xi$ & $-3\xi\cos\xi - (\xi^2-3)\sin\xi$  \\
         $j_2(\xi)$ & $\frac{3(\xi\cos\xi - \sin\xi) + \xi^2\Si(\xi)}{2\xi^2}$  & $\cos\xi - \frac{3\sin\xi}{\xi}$ & $\xi\cos\xi - 4\sin\xi + 3\,\Si(\xi)$ & $(\xi^2-8)\cos\xi - 5\xi\sin\xi$ \\
         $j''_0(\xi)$ & $\frac{\xi\cos\xi - \sin\xi}{\xi^2}$  & $\cos\xi - \frac{2\sin\xi}{\xi}$ & $\xi\cos\xi - 3\sin\xi + 2\,\Si(\xi)$ &  $(\xi^2-6)\cos\xi - 4\xi\sin\xi$ \\
         $j''_1(\xi)$ & $\frac{2\xi\cos\xi + (\xi^2-2)\sin\xi}{\xi^3}$ & $\frac{3(\xi\cos\xi-\sin\xi)}{\xi^2} + \sin\xi $ & $4\cos\xi + \frac{(\xi^2-6)\sin\xi}{\xi}$  & $5\xi\cos\xi + (\xi^2-11)\sin\xi + 6\Si(\xi)$  \\
         $j''_2(\xi)$ & $\frac{-\xi(\xi^2-9)\cos\xi + (4\xi^2-9)\sin\xi}{\xi^4}$ & $\frac{-\xi (\xi^2-12)\cos\xi + (5\xi^2-12)\sin\xi}{\xi^3}$ & $\frac{-\xi(\xi^2-18)\cos\xi + 6(\xi^2-3)\sin\xi}{\xi^2}+\Si(\xi)$ &  $-(\xi^2-24)\cos\xi + \frac{(7\xi^2-36)\sin\xi}{\xi}$ \\
    \end{tabular}
    \caption{The indefinite integrals of the kernel functions $K(\xi)$ multiplied by $1$, $\xi$, $\xi^2$, and $\xi^3$.}
    \label{tab:kernel-integrals}
\end{table*}

We calculate $\hat{K}(\omega)$ analytically as reported in Table~\ref{tab:kernel-fourier-transform-short-interval}. On the other hand, the integrals for $\hat{G}(\omega)$ given in Eq.~\eqref{eq:Ghat} are evaluated as sums. Naive integration using fast Fourier transforms leads to numerical imprecision and boundary effects (cf., Ref.~\cite{Press:2007ipz}), unless the FFT is performed on a very fine grid, hindering the performance of the calculation. Instead, we use an $N$th-order Gauss-Legendre (GL) quadrature using non-uniform fast Fourier transforms (NUFFTs),
\begin{equation}
    \hat G(\omega) = \sum_{j=1}^N w_j^{\rm GL} G(\xi_j^{\rm GL}) e^{i \omega \xi_j^{\rm GL}},
\label{eqn:forwardGL}
\end{equation}
where $w_j^{\rm GL}$ are the GL weights and $\xi_j^{\rm GL}$ the abscissas. The values of $\omega$ at which we evaluate $\hat{G}(\omega)$ and $\hat{K}(\omega)$ will be discussed below. 

There are still, however, two bottlenecks to carrying this out efficiently.\footnote{Alternatively, the convolution could be evaluated directly. While the rapid oscillations of the kernel functions make this computation expensive, it is possible by approximating $G(\xi)$ using linear interpolation over $N$ points $\xi_i\in[0,\xi_0]$. This reduces the problem to calculating integrals of the kernel functions multiplied by linear functions in $\xi$ over each subinterval $[\xi_i,\xi_{i+1}]$, which can be evaluated analytically. This direct-convolution approach avoids the ringing and aliasing artifacts intrinsic to FFT-based methods. However, we found that, given the resolution in $\xi_i$ required for comparable accuracy, evaluating the convolutions in Fourier space using the numerical schemes described later in this section remains more efficient.} First, the sharp cutoff of $G(\xi)$ at $\xi=\xi_0$ leads to ringing in Fourier space---i.e., the excitation of high-$\omega$ modes with amplitudes that fall off as $1/\omega$. These modes have nothing to do with the fairly smoothly varying source function $G(\xi)$.  The discontinuity in $G'(\xi)$ leads to ringing as well, which falls off as $1/\omega^2$.

Second, the source functions are smoothly varying for $k\tau \ll 1$ and for $ k\tau \gg 1$, but the timescales of the variation differ by up to three orders of magnitude.  The late-time behavior can in principle be resolved by a small number $N$ of Fourier modes (say up to $\omega_{\rm max} \simeq \pi N/\xi_0$ in the Fourier expansion). Similarly, the early-time behavior should also be resolved by a small number $N$ of modes, but up to a much larger value $\omega_{\rm max} \simeq \pi N/\xi_\star$, where $\xi_\star(<\xi_0)$ is what separates those early- and late-time behavior. If we want to resolve both the late-time behaviors (which requires low-$\omega$ modes) and the early-time behavior (which requires high-$\omega$ modes) together, we need 1000 times the number of modes we would need for either alone.

To deal with these two issues, we re-write the source function as $G(\xi) \equiv G_0(\xi) + G_1(\xi) + G_2(\xi)$, where
\begin{equation}
     G_0(\xi) \equiv \begin{cases} a +b \xi +c \xi^2 +d\xi^3& {\rm for} \quad 0 \leq\xi<\xi_\star, \\
     e +f \xi + g\xi^2 +h \xi^3 & {\rm for} \quad
     \xi_\star\leq\xi\leq\xi_0, \end{cases}
     \label{eq:G0}
\end{equation}
\begin{equation}
     G_1(\xi) \equiv \begin{cases} G(\xi) - G_0(\xi) & {\rm for} \quad 0 \leq\xi<\xi_\star, \\
     0 & {\rm otherwise},
     \end{cases}
     \label{eq:G1}
\end{equation}
\begin{equation}
     G_2(\xi) \equiv \begin{cases} G(\xi) - G_0(\xi) & {\rm for} \quad \xi_\star \leq\xi\leq\xi_0, \\
     0 & {\rm otherwise}.
     \end{cases}
     \label{eq:G2}
\end{equation}
The coefficients $a$, $b$, $c$, $d$, $e$, $f$, $g$, and $h$ for $G_0(\xi)$ in Eq.~\eqref{eq:G0} are chosen such that $G_1(\xi)$ and its first derivative $G_1'(\xi)$ vanish at its boundaries $\xi=0$ and $\xi=\xi_\star$, and $G_2(\xi)$ and $G_2'(\xi)$ both vanish at its boundaries $\xi=\xi_\star$ and $\xi=\xi_0$ as well.  As a result, the $1/\omega$ and $1/\omega^2$ ringing from the boundaries are suppressed for the functions $G_1(\xi)$ and $G_2(\xi)$ making them well represented by their Fourier transforms integrated over the range $-\omega_{\rm max} <\omega< \omega_{\rm max}$ with reasonably small values of $\omega_{\rm max}$.

The convolution $ I(\xi)= \int_0^\xi\, d\xi'\, G(\xi') K(\xi-\xi')$ is then evaluated in three steps as follows:  First, we define, after a change of integration variable from $\xi'$ to $\xi-\xi'$,
\begin{eqnarray}
    I_0(\xi) &=& \int_0^\xi\, d\xi'\, G_0(\xi-\xi') K(\xi').
\end{eqnarray}
Since $G_0(\xi)$ is a cubic polynomial of $\xi$ [Eq.~\eqref{eq:G0}], this convolution can be evaluated analytically using the formulas listed in Table \ref{tab:kernel-integrals}.

For the second convolution $I_1(\xi) = \int_0^\xi d\xi'\;G_1(\xi')K(\xi-\xi')$, we first calculate
\begin{equation}
    \hat G_1(\omega_l) = \sum_{j=1}^{N_{1}^{(\xi)}} w_j^{\rm GL} G_1(\xi_j^{\rm GL}) e^{i \omega_l \xi_j^{\rm GL}},
    \label{eq:G1-hat}
\end{equation}
with $\xi_j^{\rm GL}$ the $N_{1}^{(\xi)}$-th order GL abscissas spaced within $0\leq\xi\leq\xi_\star$ and corresponding weights $w_j^{\rm GL}$. By default we set $N_{1}^{(\xi)}=201$.  While $G_1(\xi)$ is restricted in $0\leq\xi\leq\xi_\star$ as we need to calculate $I_1(\xi)$ for the whole range of $\xi$, $0\leq\xi\leq\xi_0$, we need a smaller minimum Fourier frequency $\pi/\xi_0$, instead of $\pi/\xi_\star$. For this, we define a set of $N_{1}^{(\omega)} \equiv \lceil N_{1}^{(\xi)} (\xi_0/\xi_\star)\rceil$ uniform frequencies $\omega_l$'s with $\Delta \omega_l=(\pi/\xi_\star)(N_{1}^{(\xi)} / N_{1}^{(\omega)})$ centered at $\omega_l=0$, where $\lceil x \rceil$ is the smallest integer larger than $x$. Then, with the kernel $\hat K(\omega)$ obtained from its analytic expression in Table \ref{tab:kernel-fourier-transform-short-interval}, we calculate $I_1(\xi)$ as
\begin{eqnarray}
     I_1(\xi) &=& \int_{-\infty}^\infty\, d\omega \,\hat G_1(\omega) \hat K(\omega) e^{-i\omega\xi}\nonumber\\
       &\approx& \sum_{l=1}^{N_{1}^{(\omega)}} \Delta \omega_l \,e^{-i \xi \omega_l} \hat K(\omega_l) \hat G_1(\omega_l).
\label{eqn:I1ofxi}       
\end{eqnarray}

For the last convolution $I_2(\xi)= \int_0^\xi d\xi'\;G_2(\xi')K(\xi-\xi')$, we calculate
\begin{equation}
    \hat G_2(\omega_l) = \sum_{j=1}^{N_{2}^{(\xi)}} w_j^{\rm GL} G_2(\xi_j^{\rm GL}) e^{i \omega_l \xi_j^{\rm GL}},
    \label{eq:G2-hat}
\end{equation}
with $\xi_j^{\rm GL}$ the $N_{2}^{(\xi)}$-th order GL abscissas spaced in $0\leq\xi\leq\xi_0$, at $N_{2}^{(\omega)} \equiv 10N_{2}^{(\xi)}+1$ uniform frequencies with $\Delta \omega_l= \pi/(10\xi_0)$ centered at $\omega_l=0$. By default we set $N_{2}^{(\xi)}$ to be $k$-dependent with five $k$-bins defined:
\begin{equation}
N_2^{(\xi)} \equiv \begin{cases}1000 &{\rm if\;} k\leq0.1\;{\rm Mpc}^{-1},\\
5000 &{\rm if\;} 0.1\;{\rm Mpc}^{-1}< k\leq1\;{\rm Mpc}^{-1},\\
25000 &{\rm if\;} 1\;{\rm Mpc}^{-1}< k\leq5\;{\rm Mpc}^{-1},\\
40000 &{\rm if\;} 5\;{\rm Mpc}^{-1}< k\leq50\;{\rm Mpc}^{-1},\\
80000 &{\rm otherwise}.
\label{eq:N2-xi}
\end{cases}
\end{equation}
Note that this $k$-dependent definition of $N_2^{(\xi)}$ can be further optimized, with more than five $k$-bins. All the $N_2^{(\xi)}$-th order GL abscissas and weights are pre-calculated and tabulated in \texttt{/GL} folder. 
The kernel $\hat K(\omega_l)$ is again obtained from its analytic expression in Table \ref{tab:kernel-fourier-transform-short-interval}, as those for $I_1(\xi)$. Then $I_2(\xi)$ can be obtained by
\begin{equation}
    I_2(\xi) \approx \sum_{l=1}^{N_{2}^{(\omega)}} \Delta w_l \, e^{-i \xi \omega_l} \hat K(\omega_l) \hat G_2(\omega_l).
\end{equation}
Note that $I_2(\xi)$ vanishes for $\xi<\xi_\star$ simply because $G_2(\xi)=0$ within that range, hence we calculate $I_2(\xi)$ only for $\xi_\star\leq\xi\leq\xi_0$. Finally, then, we have
\begin{equation}
     I(\xi) = \begin{cases} I_0(\xi) + I_1(\xi)& {\rm for} \quad 0 \leq\xi<\xi_\star \\
      I_0(\xi) + I_1(\xi) + I_2(\xi) & {\rm for} \quad \xi_\star \leq\xi\leq\xi_0.    \end{cases}
\end{equation}
We implement this into the Boltzmann solver \texttt{CLASS} by mainly modifying \texttt{source/perturbations.c}, with some useful auxiliary functions for our implementation provided in \texttt{tools/ncdmfft\_tools.c}. To be specific, we first define a set of conformal times $\tau_{\rm NCDM}$'s, half of which are evenly-spaced in log-scale for $\tau_i \leq \tau_{\rm NCDM} <10^3\;{\rm Mpc}$ and half of which are evenly-spaced in linear-scale for $10^3\;{\rm Mpc}\leq\tau_{\rm NCDM}\leq\tau_0$.\footnote{Here $\tau_i$ is when the initial conditions for cosmological perturbations are set in \texttt{CLASS}. While this split for different $\tau$-sampling scales does not perfectly coincide with our choice of $\xi_\star$ given in Section \ref{sec:xi1}, it still allows to capture behaviors of perturbations in two different epochs properly. In addition, it is computationally more effective to have a global $\tau$-grid to store the metric perturbations for all $k$ and $q$.} At these $\tau_{\rm NCDM}$'s, we tabulate the metric-perturbation variables $\dot{h}$ and $\dot{\eta}$. These metric-perturbation variables are interpolated onto the rescaled GL abscissas $\xi_j^{\rm GL}$(k,q)'s in Eq.\;\eqref{eq:G1-hat} and \eqref{eq:G2-hat}, and used to perform forward Fourier transforms. The set of $\tau_{\rm NCDM}$'s defined earlier to tabulate $\dot{h}$ and $\dot{\eta}$ defines the output grids of $\xi(k,q,\tau_{\rm NCDM})$ as well, on which we perform the backward Fourier transforms obtaining the integral solutions. We later interpolate those solutions in $\tau$ whenever needed to compute the matter-density, pressure perturbations, peculiar velocity, and anisotropic stress. 

To compute the forward and backward integrations with discrete input/output grids, we perform type-3 (non-uniform to non-uniform) NUFFTs using \texttt{FINUFFT} (Flatiron Institute Nonuniform Fast Fourier Transform) \cite{barnett2019parallelnonuniformfastfourier,barnett2020aliasingerrorexpbetasqrt1z2}.\footnote{\href{https://finufft.readthedocs.io/en/latest/}{https://finufft.readthedocs.io/en/latest/}}  Note that all the forward Fourier transforms performed with \texttt{FINUFFT} are vectorized for three $\ell$'s and for two types of terms [Eq.~\eqref{eq:GA} and \eqref{eq:GB}], and the backward transforms are done only once after adding the contributions from both types of terms, hence vectorized only for three $\ell$'s.

\subsection{Choice of $\xi_\star$}
\label{sec:xi1}

As pointed out earlier, the source functions vary smoothly in both the super-/near-horizon ($k\tau \lesssim 1$) and sub-horizon ($k\tau \gg 1$) regimes, but the timescales of variation differ significantly. This behavior arises because each mode evolves slowly when outside the horizon, but begins to evolve more rapidly due to causal physics once it enters the horizon. When a mode is well inside the horizon, the metric perturbations have already responded to the fluid perturbations and settle into a quasi-static configuration, leading to slower evolution once again.

To accurately and effectively capture the behavior of the source functions in both regimes, we define a transition parameter $\xi_\star^{\rm default}=100$, which separates the two regimes by requiring the horizon scale to be two orders of magnitude larger than the scale of the mode.

There is one caveat, though. Low-momentum particles have correspondingly low $\xi_0$, which can become comparable to or even smaller than $\xi_\star^{\rm default}$. In such cases, however, the source term $G(\xi)$ involving both $\dot{\eta}$ and $\dot{h}$ [Eq.~\eqref{eq:GB}] increases rapidly with $\xi$ at late times as $d\chi_q/d\tau \ll 1$. Hence, separating out this rapidly growing evolution helps maintain numerical stability and accuracy. For this reason, we define 
\begin{equation}
\xi_\star\equiv {\rm max}(\xi_\star^{\rm default}, \xi_0/2).
\end{equation}
That is, by $\xi_\star$ we separate super-/near-horizon and sub-horizon regimes for relativistic particles at $\xi_\star^{\rm default}$, and relatively slowly varying behavior in early time and rapidly varying behavior in late times for non-relativistic particles at $\xi_0/2$.

\subsection{Big-$q$ approximation}
\label{sec:big-q}

Our calculation involves values of $q$ and $k$ spanning a broad range, and we have also separated the calculation into super-/near-horizon and sub-horizon regimes using $\xi_\star$ for particles still relativistic at $\xi_\star$. In the super-/near-horizon regime, the momentum dependence of the source function $G_1$, given by $G_1\propto\frac{\epsilon}{q}\frac{d\ln f_0}{d\ln q}$, can be factored out when $q$ is sufficiently large such that the particle remains relativistic ($\epsilon\approx q$) up to $\xi=\xi_\star$. In this case, the convolution with $G_1$ only needs to be performed once for the largest momentum $q_{\rm max}$ for a given $k$-mode. Afterward, the result can be easily rescaled and interpolated in $\xi$ for the remaining momentum bins. Although $\xi$ depends on $q$ at late times (since $\epsilon=\sqrt{q^2+a^2m^2/T_0^2}$), making the relation between $\xi$ and $\tau$ different for each $q$, this approximation remains valid because the convolution is expressed purely in terms of $\xi$, and the $q$-dependence is implicitly absorbed in it.

Since the final convolution with $G_1$ depends on $\xi_\star$, this approximation can only be applied to $q$-bins that share the same $\xi_\star$. Recall that we define $\xi_\star\equiv {\rm min}(\xi_\star^{\rm default}, \xi_0/2)$, so $\xi_\star$ generally varies with $q$ and $k$. To identify momentum bins for which the particle remains relativistic until $\tau(\xi_\star)$, we impose the following criterion:
\begin{equation}
\Bigg|\frac{\xi_\star^{\rm default}/k - \tau(\xi_\star)}{\tau(\xi_\star)}\Bigg|< \texttt{ppr->ncdmfft\_bigq\_approx\_tol},
\end{equation}
where the default tolerance is set as $\texttt{ppr->ncdmfft\_bigq\_approx\_tol}=10^{-3}$. Because the convolution with $G_1$ requires a finely sampled set of FFT frequencies contributing significantly to the total runtime, this approximation is numerically very efficient. It is especially beneficial when working with large $k_{\rm max}$ values or a high-resolution momentum grid.

\subsection{Convergence of iterations}
\label{sec:convergence}

\begin{figure*}[t!]
\includegraphics[width = 2\columnwidth,trim= 20 20 20 0]{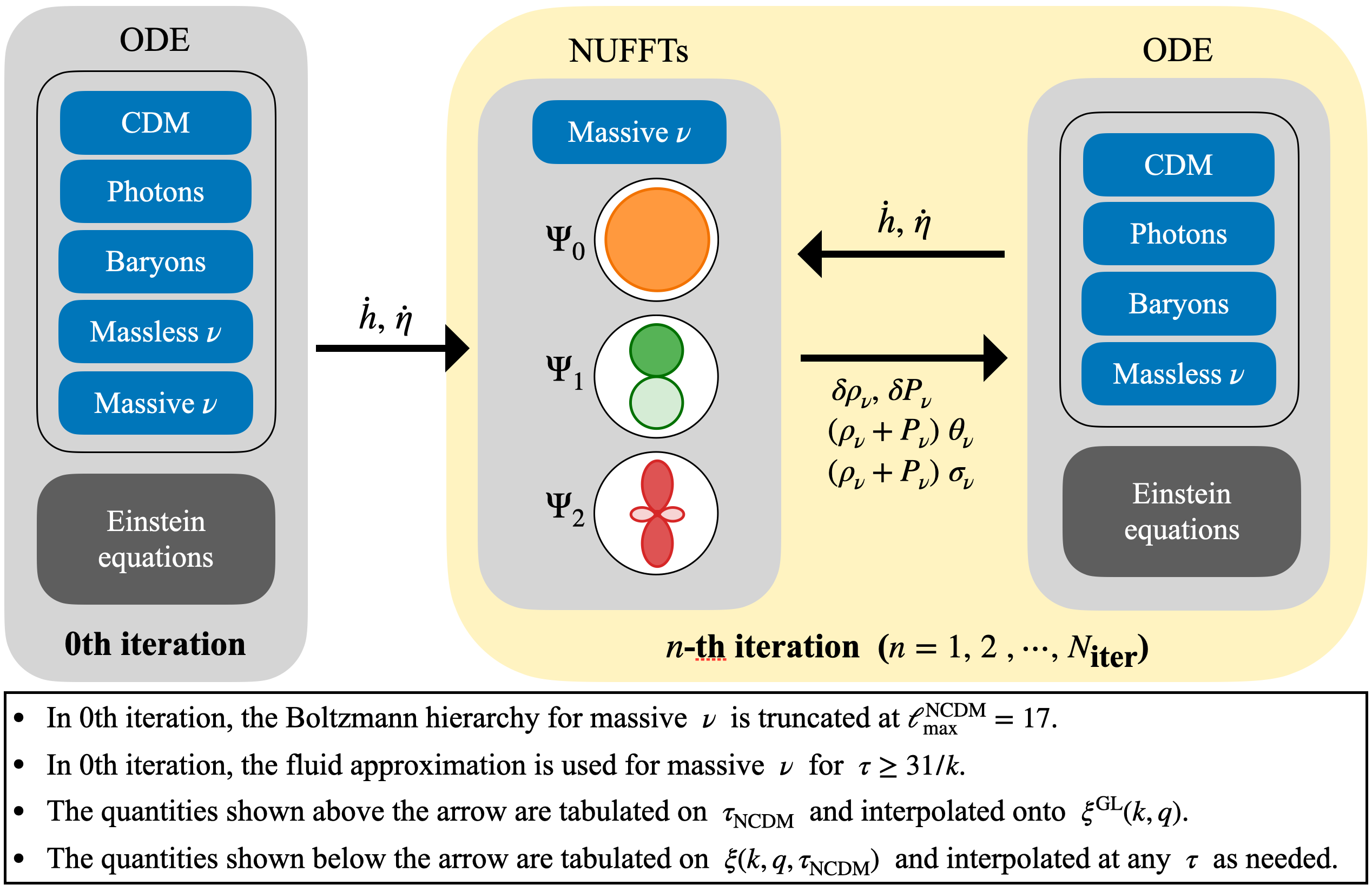}
\caption{Schematic diagram describing our iterative approach with the Boltzmann hierarchy for massive neutrino replaced by the integral solutions after 0th iteration.}
\label{fig:diagram}
\end{figure*}

\begin{figure*}[ht!]
\includegraphics[width = \linewidth,trim= 0 0 0 0]{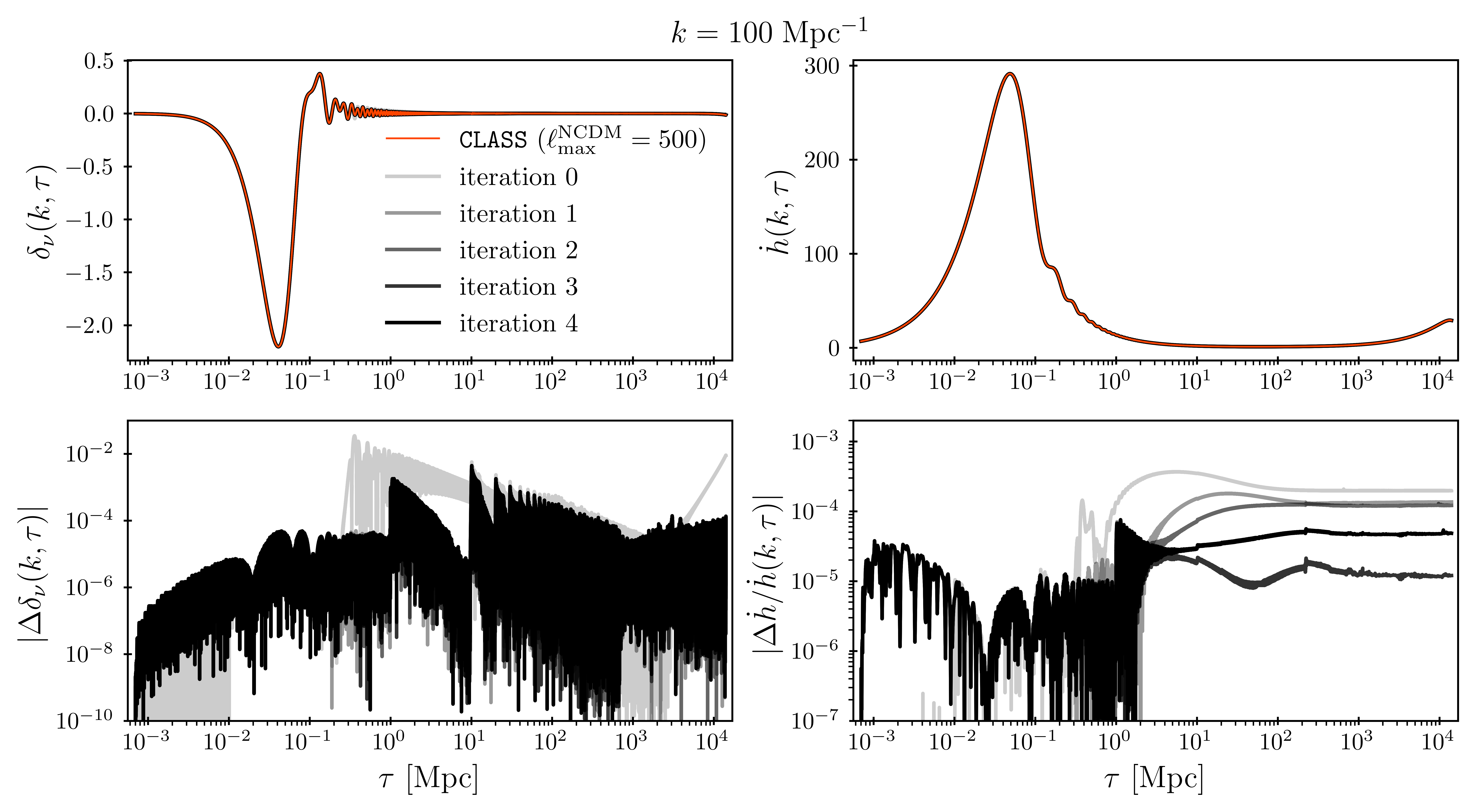}
\includegraphics[width = \linewidth,trim= 0 10 0 0]{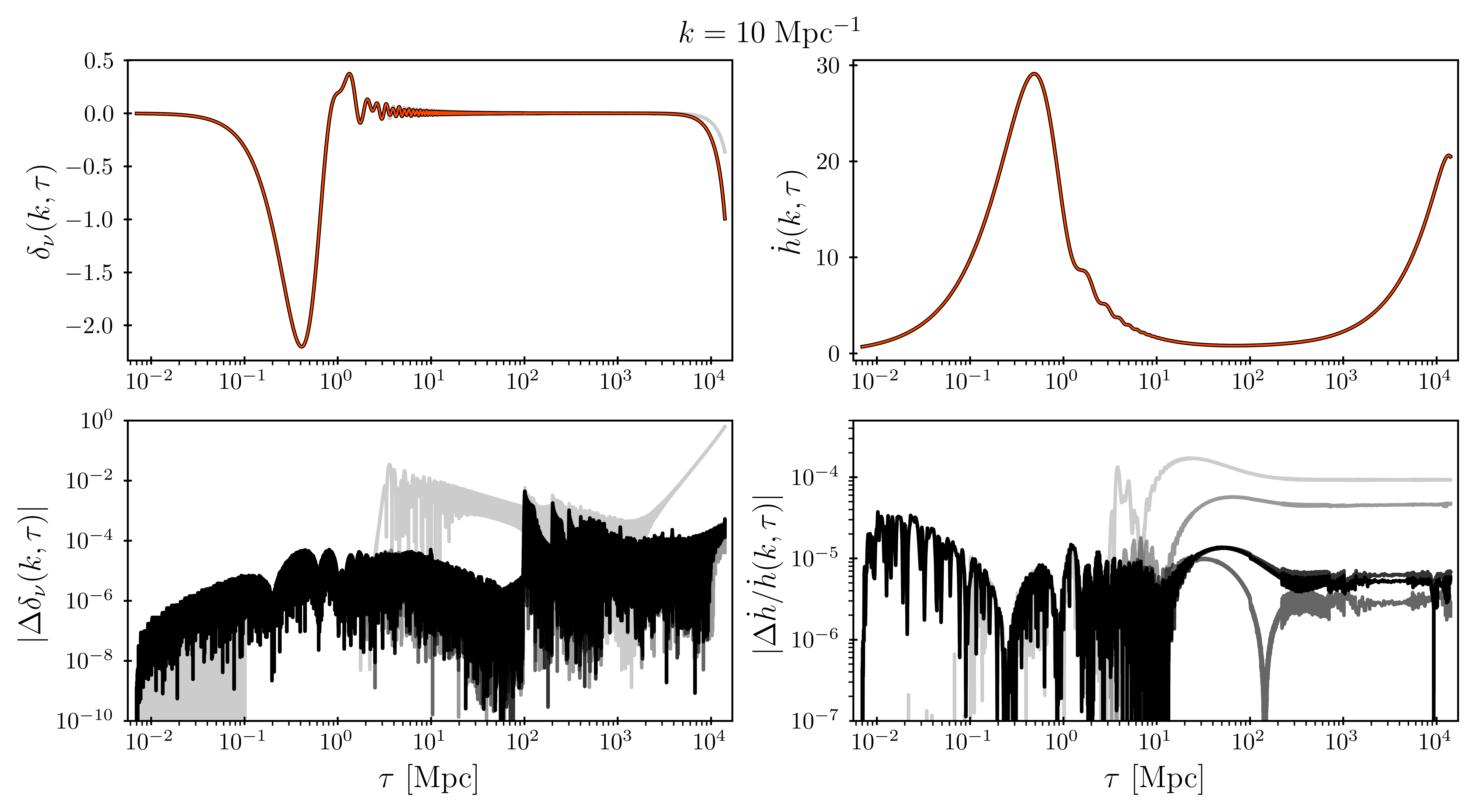}
\caption{The numerical solutions $\delta_\nu$ (left) and $\dot{h}$ (right) from the original \texttt{CLASS} (orange) and our implementations \texttt{CLASSIER} (light grey to black) with up to four iterations, for $k=100\;{\rm Mpc}^{-1}$ mode (upper plots) and $k=10\;{\rm Mpc}^{-1}$ mode (lower plots). Bottom panel of each plot shows the absolute differences for $\delta_\nu$ and fractional differences for $\dot{h}$ from each iteration result, with respect to the standard \texttt{CLASS} calculations with $\ell_{\rm max}^{\rm NCDM}=500$, which demonstrates the rapid convergence of our results. Note that the intermediate iterations oscillate around the final converged solution. A seemingly large discrepancy in $\delta_\nu$ for $k=100\;{\rm Mpc}^{-1}$ mode at $\tau=1$ Mpc (the bottom panel of upper left plot) is due to the numerical error in $I_2(\xi)$ at $\xi=\xi_\star$, the convolution involving $G_2(\xi)$, and is only $\mathcal{O}(1\%)$ error. The repeatedly appearing errors in $\delta_\nu$ starting at $\tau \sim 10$ Mpc for $k=100\;{\rm Mpc}^{-1}$ and at $\tau\sim10^2$ Mpc for $k=10^{-1}\;{\rm Mpc}^{-1}$ are due to the numerical errors in \texttt{CLASS} calculations due to the truncation of the Boltzmann hierarchy (see the second inset plot of Fig.~\ref{fig:delta_nu-k100}). The times when the 0th iteration result start deviating from \texttt{CLASS} results in each plot is when the fluid approximation for massive neutrino turns on for corresponding $k$-mode.}
\label{fig:converg-k100-k10}
\end{figure*}

\begin{figure*}[t!]
\includegraphics[width = \linewidth,trim= 0 0 0 0]{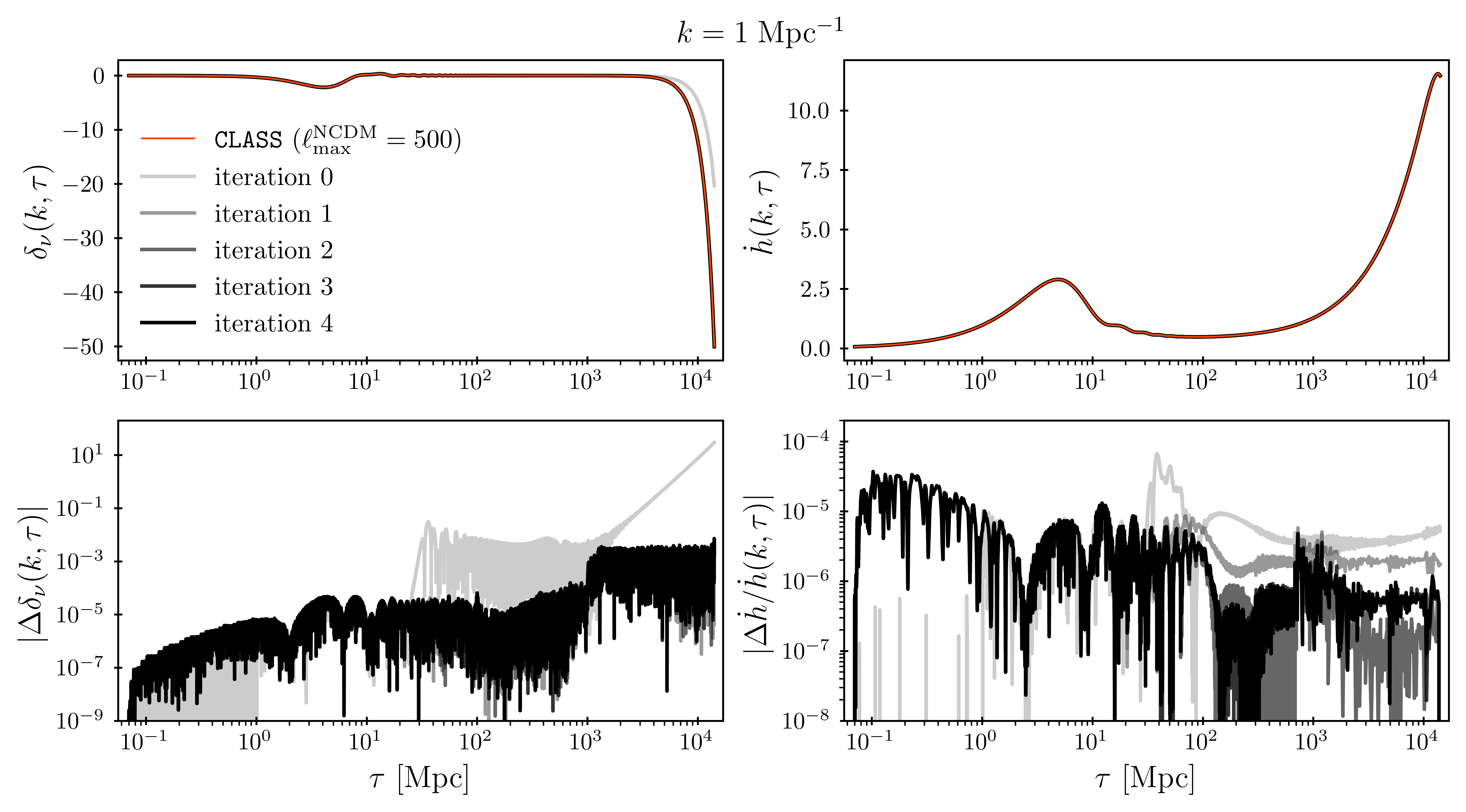}
\includegraphics[width = \linewidth,trim= 0 10 0 0]{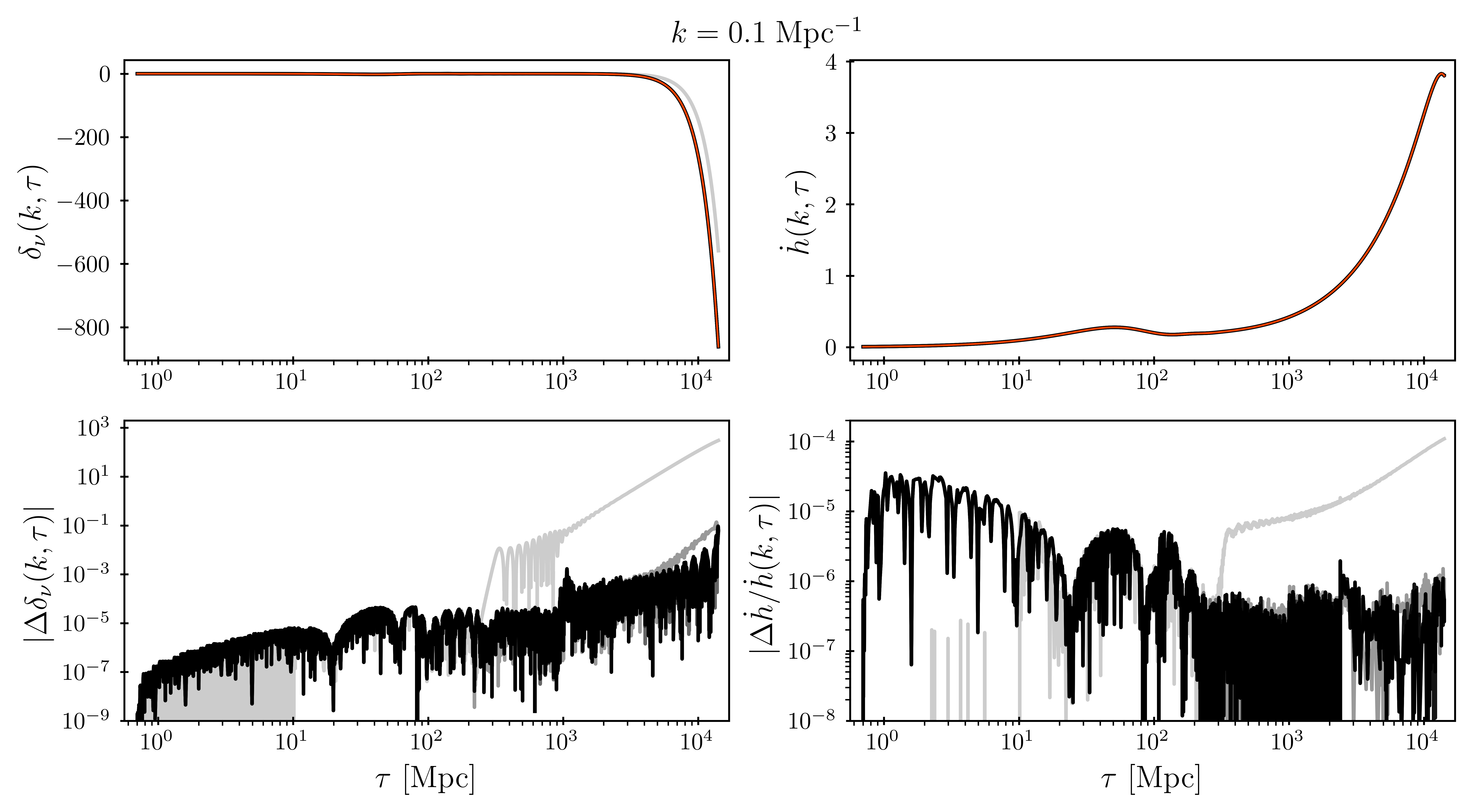}
\caption{Same as Fig.~\ref{fig:converg-k100-k10}, but for $k=1\;{\rm Mpc}^{-1}$ (upper plots) and $0.1\;{\rm Mpc}^{-1}$ (lower plots).}
\label{fig:converg-k1-k0.1}
\end{figure*}

We demonstrate the rapid convergence of our iterative scheme when applied to massive neutrinos as illustrated in Fig.~\ref{fig:diagram}, in this subsection. Figures~\ref{fig:converg-k100-k10} and \ref{fig:converg-k1-k0.1} illustrate the convergence behavior of our solutions for wavenumbers $k=100,\;10,\;1,\;0.1\;{\rm Mpc}^{-1}$. We show $\delta_\nu$ and $\dot{h}$ as representative examples; the convergence of other perturbation variables is similarly robust. In the bottom panels, we plot the absolute difference in $\delta_\nu$ and the absolute fractional difference in $\dot{h}$ relative to the standard \texttt{CLASS} results with $\ell_{\rm max}^{\rm NCDM}=500$, a setting that ensures convergence of the matter power spectrum today.
 
While the matter power spectrum converges at $\ell_{\rm max}^{\rm NCDM} \sim 500$, the individual perturbation variables still inherit numerical errors at early times due to the truncation of the Boltzmann hierarchy, errors that are absent in our method. We defer a more detailed comparison of the two approaches to Section~\ref{sec:comparison}, and focus here solely on the convergence of our iterative method.
 
For clarity, we show results up to the 4th iteration. Each subsequent iteration fully overlaps with the 4th iteration, demonstrating clear convergence as shown. Our results may not converge to the standard \texttt{CLASS} solutions, but the differences are negligible. The bottom panels confirm that the differences between successive iterations decrease steadily, indicating numerical convergence. For most practical applications, enough precision is achieved after at most 2 iterations.

\section{Comparison with Standard Boltzmann Hierarchy approach}
\label{sec:comparison}

\begin{figure}[t!]
\includegraphics[width = .95\columnwidth,trim= 20 20 20 0]{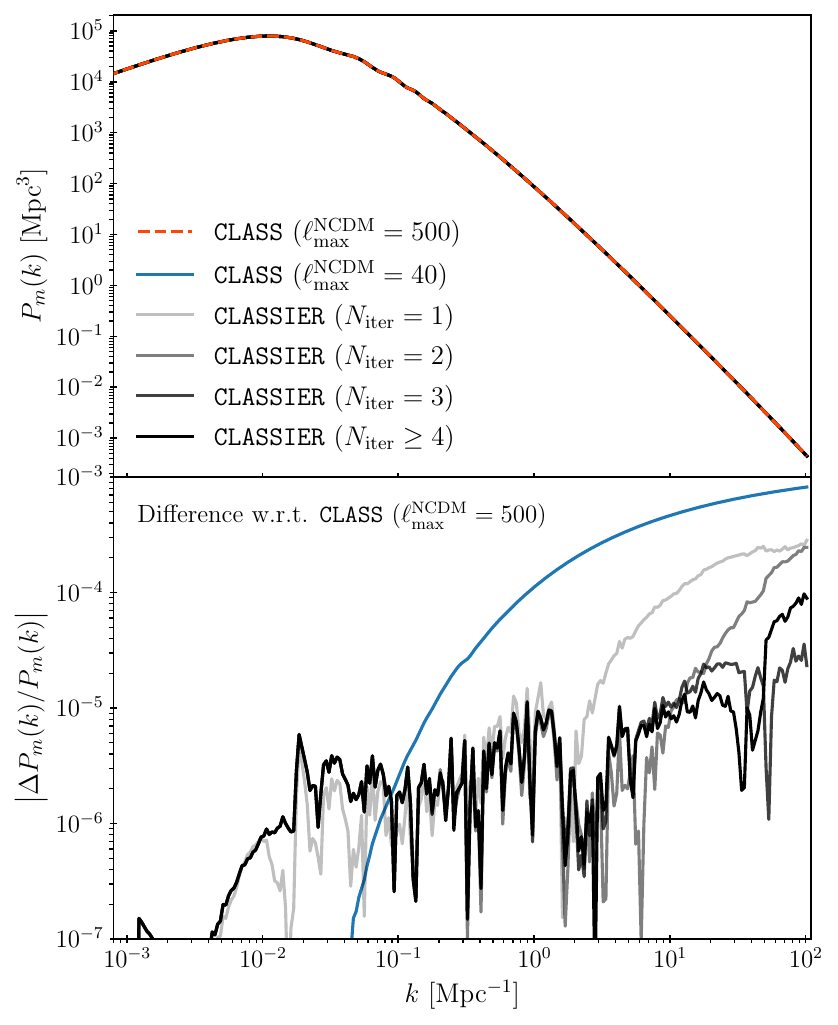}
\caption{Top: Matter power spectrum today from three different calculations: our implementation \texttt{CLASSIER} (light grey to black), the standard Boltzmann hierarchy approach \texttt{CLASS} truncated at $\ell_{\rm max}^{\rm NCDM}=40$ (blue) and $\ell_{\rm max}^{\rm NCDM}=500$ (orange, dashed) without the fluid approximation for massive neutrino. Bottom: the fractional differences of $P_m(k)$ from the truncated Boltzmann hierarchy with respect to that from our integral-equation approach.}
\label{fig:Pm}
\end{figure}

\begin{figure*}[t!]
\includegraphics[width = \linewidth,trim= 0 20 0 0]{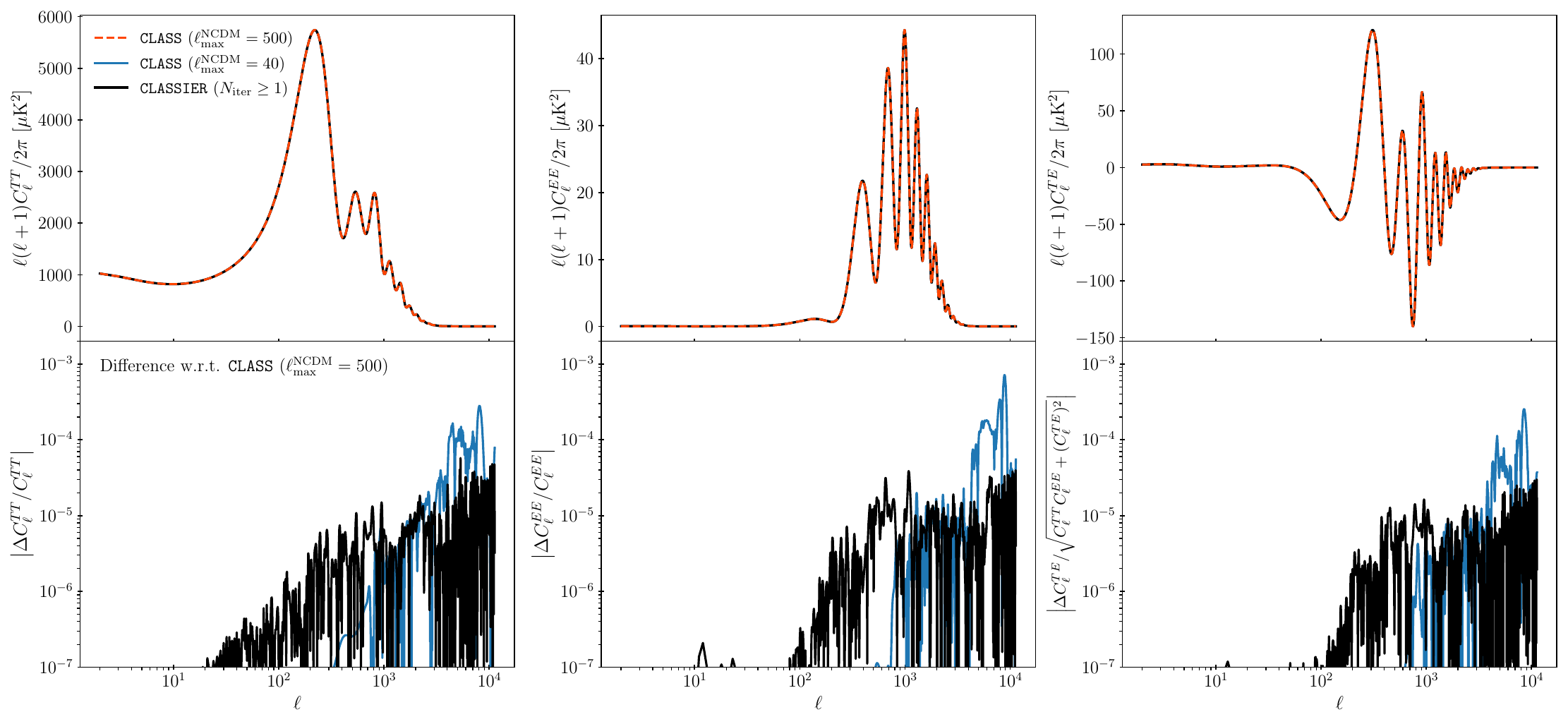}
\caption{The unlensed CMB TT (left), EE (middle), and TE (right) angular power spectra from three different calculations: our implementation \texttt{CLASSIER} (black), the standard Boltzmann hierarchy approach \texttt{CLASS} truncated at $\ell_{\rm max}^{\rm NCDM}=40$ (blue) and $\ell_{\rm max}^{\rm NCDM}=500$ (orange, dashed) without the fluid approximation for massive neutrino. The bottom panel of each plot shows the fractional differences with respect to \texttt{CLASS} ($\ell_{\rm max}^{\rm NCDM}=500$) calculations.}
\label{fig:Cell's}
\end{figure*}

In this Section, we compare our results from the integral-equation approach with those from the standard Boltzmann hierarchy approach, truncated at finite $\ell_{\rm max}^{\rm NCDM}$ and without employing the fluid approximation for massive neutrinos. For comparison, we fix all the precision parameters of $\texttt{CLASS}$ to the \texttt{CLASS}:UHP set in Ref.~\cite{Euclid:2024imf}. This set of precision parameters achieves $<\mathcal{O}(0.1\%)$ errors in the matter power spectrum today with massive neutrinos without the fluid approximation up to $k \sim 10\;{\rm Mpc}^{-1}$, with respect to high precision \texttt{CAMB} calculation. While $\ell_{\rm max}^{\rm NCDM}=40$ defined in \texttt{CLASS}:UHP already gives quite accurate results, we also compare to a highly converged solution using $\ell_{\rm max}^{\rm NCDM}=500$.

We compare the matter power spectrum today in Fig.~\ref{fig:Pm}. The bottom panel shows the fractional differences. Remarkably, even a single iteration (labeled as $N_{\rm iter}=1$ in the figures) yields highly accurate results when compared with the standard \texttt{CLASS} calculations with $\ell_{\rm max}^{\rm NCDM}=500$, with $\mathcal{O}(0.01\%)$ errors for $10\;{\rm Mpc}^{-1} \lesssim k \lesssim 100\;{\rm Mpc}^{-1}$ and less than 0.01\% errors for the rest of $k$-modes. With a few more iterations our $P_m(k)$ agrees to better than $0.01\%$ for the entire scales considered, demonstrating excellent agreement.

Let us use Figs.~\ref{fig:converg-k100-k10}, \ref{fig:converg-k1-k0.1} and ~\ref{fig:Pm} to examine the differences in the results between the two approaches. Regarding neutrino perturbations, we find first a seemingly large discrepancy in $\delta_\nu$ for $k=100\;{\rm Mpc}^{-1}$ mode at $\tau=1$ Mpc (the bottom panel of upper left plot in Fig.~\ref{fig:converg-k100-k10}). This discrepancy is due to the numerical error in $I_2(\xi)$ at $\xi=\xi_\star$, the convolution involving $G_2(\xi)$, and is only $\mathcal{O}(1\%)$ error. The same discrepancy can be seen in $\dot{h}$ as well. We checked that by increasing the number of $\xi$ points for $G_2(\xi)$ we can eliminate this discrepancy but at the cost of other adjustments for accurate late time results such as $\tau$-sampling and FFT frequency sampling. As this error is only of $\mathcal{O}(1\%)$ at very early epoch and does not affect any relevant observable, we keep the current setting. The discrepancies in $\delta_\nu$ around $\tau \sim 10^3/k$ arise from the numerical artifacts in \texttt{CLASS} calculations due to the truncation of the Boltzmann hierarchy (see the second inset plot of Fig.~\ref{fig:delta_nu-k100}). These are artificial power reflection due to truncation, where missing higher multipole moments reflect back into lower-order modes, a known artifact of truncating the Boltzmann hierarchy \cite{Ma:1995ey,class}. We emphasize that our results do not suffer from these truncation errors, as shown in the inset plots of Fig.~\ref{fig:delta_nu-k100}.

Additionally, the Fig.~\ref{fig:converg-k100-k10} and Fig.~\ref{fig:converg-k1-k0.1} show how inaccurate the fluid approximation for massive neutrinos is for neutrino density perturbations (see results for the 0th iteration).

Our implementation is robust with early Universe observables such as CMB anisotropy spectra as well. This is expected as the numerical difficulties of our method typically arise at late time, and with large $k$-modes. For completeness, we present the unlensed CMB TT, EE, and TE spectra in Fig.~\ref{fig:Cell's} with their absolute fractional differences with respect to \texttt{CLASS} ($\ell_{\rm max}^{\rm NCDM}=500$) case. We only show our $N_{\rm iter}=1$ case as all the spectra converges even with a single iteration being consistent with the standard Boltzmann hierarchy approach at $<\mathcal{O}(0.01\%)$.

Our numerical approach is not only as accurate as solving the full Boltzmann hierarchy, but it can also be much faster than the truncated Boltzmann hierarchy approach, depending on the values of $k_{\rm max}$ and $\ell_{\rm max}^{\rm NCDM}$. We present runtimes of the two approaches taken to solve the entire system under the \texttt{CLASS}:UHP precision setting, varying $k_{\rm max}$ in Table.~\ref{tab:runtime}. These results demonstrate that our approach is substantially faster, particularly as $k_{\rm max}$ and $\ell_{\rm max}^{\rm NCDM}$ increase, while our method retains excellent accuracy.

\begin{table}[!ht]
  \centering
  \begin{tabular}{c||c|c|c|c}
  $k_{\rm max}$ &  \multicolumn{2}{c|}{\texttt{CLASSIER}} & \multicolumn{2}{c}{\texttt{CLASS}}     \\ 
  $({\rm Mpc}^{-1})$ & ~$N_{\rm iter}=1$~ & ~$N_{\rm iter}=2$~ & ~$\ell_{\rm max}^{\rm ncdm}=40$~ & ~$\ell_{\rm max}^{\rm ncdm}=500$~ \\
  \hline
  \hline
  $1$  & 3.2s & 5.7s & 1.8s  & 16s \\
    \hline
  $10$ & 13s & 24s  & 16s  & 2m 30s  \\
    \hline
  $50$ & 30s & 53s  & 1m 20s & 12m 40s \\
    \hline
  $100$  & 47s & 1m 22s  & 2m 40s & 25m
  \end{tabular}
  \caption{Runtimes (on four threads of 16-inch MacBook Pro -- M4 chip, Nov 2024 release) for solving the full system under the \texttt{CLASS}:UHP precision setting, for different $k_{\rm max}$. Single-thread runtimes are roughly four times longer.}
\label{tab:runtime}
\end{table}

\section{A test with a toy phase-space density distribution for non-cold relics}

\begin{figure}[!ht]
\includegraphics[width = .95\columnwidth,trim= 20 20 20 0]{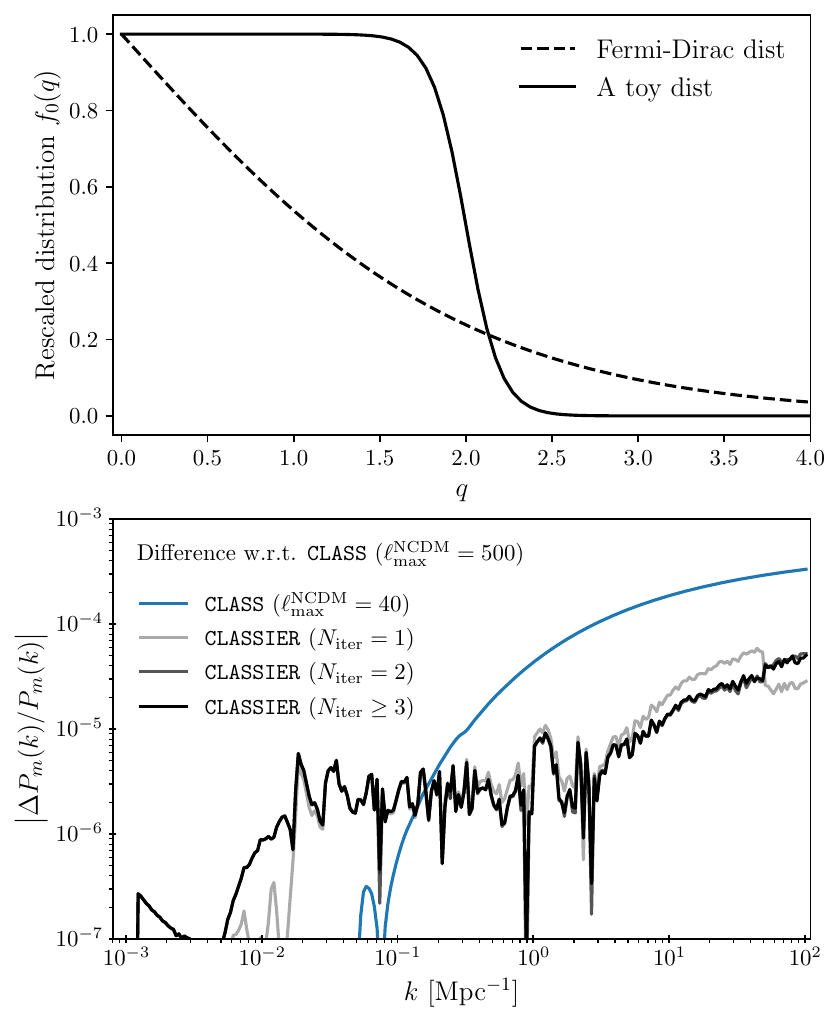}
\caption{Top: Rescaled background phase-space density distributions: the usual Fermi-Dirac for massive neutrino (dashed) and a toy distribution used for test (solid). Bottom: The fractional difference of matter power spectrum today obtained with a toy phase-space density distribution assumed for non-cold relics. The differences are given with respect to that of \texttt{CLASS} ($\ell_{\rm max}^{\rm NCDM}=500$) calculation.}
\label{fig:Pm-nonFD}
\end{figure}

So far, we have only considered the standard non-cold relics, massive neutrinos. In this Section, we present a simple test for the validity of our approach with non-cold relics other than the standard massive neutrino. To this end, instead of one massive neutrino species with Fermi-Dirac distribution, we consider non-cold relics with a simple toy phase-space density distribution, but accounting for the same energy density as $m_\nu=0.06\;{\rm eV}$ massive neutrino, which is $\omega_{\rm NCDM}= 6.44\times10^{-4}$. We show the toy distribution we assumed, and the fractional differences of resulting matter power spectrum compared to the standard truncated Boltzmann hierarchy approach, in Fig.~\ref{fig:Pm-nonFD}. This test demonstrates that our IE approach can be a useful tool for studying non-cold relics.\\

Altogether, these comparisons and the test confirm that our IE method offers a fast and highly accurate alternative to the traditional Boltzmann hierarchy, particularly for applications requiring high $k$-resolution and precise treatment of non-cold species.

\section{Conclusion}
\label{sec:conclusion}

We have introduced and implemented a novel approach to evolving perturbations of non-cold dark matter (NCDM) species, based on an exact integral-equation formulation. This method circumvents the need to solve the full Boltzmann hierarchy by instead evaluating a set of convolution integrals. Our approach achieves agreement with the full Boltzmann solution with difference smaller than $0.01\%$ in the matter power spectrum at $z=0$, up to $k=100\;{\rm Mpc}^{-1}$, while substantially reducing computational cost. In particular, it outperforms the truncated Boltzmann hierarchy at high $k$ values, both in speed and accuracy, and avoids the numerical artifacts associated with the truncation of Boltzmann hierarchy at finite $\ell_{\rm max}^{\rm NCDM}$. Moreover, our method requires no model-specific fluid approximations and is applicable to arbitrary NCDM species, including massive neutrinos and other non-cold relics with non-standard phase-space distributions. While we considered massive neutrino for demonstration where the 0th iteration involves a fluid approximation, this step is not required for more general non-cold relics and does not reflect a limitation of the integral-equation method itself. Further, given the accuracy of the fluid approximation for massive neutrinos, we expect that the relative effect of the miscalibration will not affect the efficiency of the convergence when applied to other non-cold relics.

By enabling fast and accurate treatment of non-cold species, our method is particularly well suited for applications involving high-$k$ observables, such as future large-scale structure surveys, studies of the small-scale secondary anisotropies of the CMB, or scenarios with non-standard thermal histories. Possible applications of our approach include, but are not limited to, decaying dark matter \cite{Aoyama:2014tga,FrancoAbellan:2021sxk}, warm dark matter \cite{DEramo:2020gpr,Banerjee:2025gwe}, sterile neutrinos \cite{Venumadhav:2015pla,Abazajian:2019ejt}, non-standard neutrino phase-space density \cite{Alvey:2021sji}, and $N$-naturalness models \cite{Bansal:2024afn}. Our iterative scheme also has the potential to replace the collisional Boltzmann hierarchy (e.g., neutrino self-interaction \cite{Oldengott:2017fhy} and DM-massive neutrino interactions \cite{Mosbech:2020ahp}) by treating the collision term computed in the previous iteration as an additional source. In future work, we plan to extend this approach to include vector and tensor perturbations, as well as generalizations to non-flat cosmologies.

Overall, the integral-equation method presented here provides a practical and theoretically robust alternative to the Boltzmann hierarchy, offering new opportunities for efficient precision cosmology in the era of high-precision measurements.

\section*{Acknowledgments}

We thank J.~Lesgourgues for useful discussions. This work was supported at JHU by NSF Grant No.\ 2412361, NASA ATP Grant No.\ 80NSSC24K1226, the Guggenheim Foundation, and the Templeton Foundation. NL was supported by the Horizon Fellowship from Johns Hopkins University. JLB acknowledges funding from the Ramón y Cajal Grant RYC2021-033191-I, financed by MCIN/AEI/10.13039/501100011033 and by the European Union “NextGenerationEU”/PRTR, as well as the project UC-LIME (PID2022-140670NA-I00), financed by MCIN/AEI/ 10.13039/501100011033/FEDER, UE. MK thanks the Center for Computational Astrophysics at the Flatiron Institute and the Institute for Advanced Study for hospitality.

\section*{Data Availability}

The data that support the findings of this article are openly available \cite{CLASSIER}.

\bibliography{mybib}
\end{document}